# Government as Network Catalyst: Accelerating Self-Organization in a Strategic Industry

Travis A. Whetsell
Department of Public Policy & Administration, School of International & Public Affairs,
Florida International University,
11200 SW 8th Street, Miami, Florida 33199,
travis.whetsell@fiu.edu

Michael D. Siciliano
Department of Public Administration,
University of Illinois at Chicago, Chicago, IL

Kaila G.K. Witkowski
Department of Public Policy & Administration, School of International & Public Affairs,
Florida International University, Miami, FL

Michael J. Leiblein
Fisher College of Business,
Ohio State University, Columbus, OH

**Abstract:** Governments have long standing interests in preventing market failures and enhancing innovation in strategic industries. Public policy regarding domestic technology is critical to both national security and economic prosperity. Governments often seek to enhance their global competitiveness by promoting private sector cooperative activity at the inter-organizational level. Research on network governance has illuminated the structure of boundary-spanning collaboration mainly for programs with immediate public or non-profit objectives. Far less research has examined how governments might accelerate private sector cooperation to prevent market failures or to enhance innovation. The theoretical contribution of this research is to suggest that government programs might catalyze cooperative activity by accelerating the preferential attachment mechanism inherent in social networks. We analyze the long-term effects of a government program on the strategic alliance network of 451 organizations in the high-tech semiconductor industry between 1987 and 1999, using stochastic network analysis methods for longitudinal social networks.

**Acknowledgements:** We would like to thank Caroline Wagner for discussion and comments. An earlier version of this research was presented at the American Society for Public Administration 2019 Annual Conference. Data for this article were originally collected with support by the National Science Foundation under Grant #1133043.

## 1. Introduction

Governments have long standing interests in science and technology for at least two primary reasons. First, technology is critical to the continued prosperity of nations that seek to secure a high standard of living for their citizens, fueled in part by dramatic enhancements in consumer goods, telecommunications technology, automobiles, optoelectronics, and computers (Salter and Martin 2001). Second, technology is critical to national security and the development of weapon systems, such as intercontinental ballistic missiles, unmanned aerial vehicles, and rapidly evolving applications of artificial intelligence (Mowery 1998). For these reasons, among others, nations often seek to enhance their competitiveness through science and technology (S&T) policy (Taylor 2016).

In a recent example, China has rapidly increased national research and development (R&D) spending over the past two decades, invested heavily in U.S. firms to acquire technology and intellectual property, often engaging in illicit transfers, and expanded intelligence collection on innovation targets. Together these efforts form the elements of a strategy to fuel economic growth and enhance military capabilities (Deutch 2018). This type of technology-oriented strategy is by no means new, and many developed nations invest heavily in national systems of science and technology. Nation states seeking to compete in the international system often attempt to gain advantage through scientific and technological development (Taylor 2016). The U.S. has maintained coherent national science and technology policies focusing on military strength and domestic prosperity since at least World War II (Bush 1945; Smith 2011).

A unique example from 1987 occurred when the United States intervened in the global high-technology market during a critical period of potential domestic market failure through a government sponsored non-profit research consortium, called Sematech (Browning, Beyer, and Shetler 1995; Ham, Linden, and Appleyard 1998; Browning and Shetler 2000; Carayannis and Alexander 2004). In contrast to direct subsidies, tariffs, illicit transfers from foreign competitors, and the creation of state-owned enterprises, the approach of Sematech represents a less well-known policy tool that encourages inter-organizational cooperation around shared public objectives. This article brings the organizational/administrative elements of S&T policy into focus and recontextualizes an exemplar case given new conceptual and methodological tools in network governance (Provan and Kenis 2008; Klijn and Koopenjan 2015) and social network analysis (Lusher, Koskinen, and Robins 2013; Snijders et al. 2010; Lubell et al. 2012).

The general theoretical proposition of this article is that government intervention can catalyze cooperative behavior around sets of target organizations in policy relevant inter-organizational networks by accelerating the preferential attachment process inherent in many types of networks. Preferential attachment refers to a process of self-organization in developing networks where new actors tend to form connections with already well-connected actors (Barabási and Albert 1999).[1] This mechanism presents a relatively unexplored pathway by which governments can accelerate network formation in inter-organizational networks. While the high-technology sector provides a unique case, the theoretical logic may generalize to other public policy relevant sectors with a history of, or potential for, cooperative activity. Previous research provides evidence that Sematech was a critical element in a broader industrial policy that enhanced the performance of U.S. firms (e.g. Irwin and Klenow 1996; Whetsell, Leiblein, Wagner 2019). This article takes a step back from performance effects to pose two broader research questions concerning intermediate network properties. Does cooperative activity between organizations in the high-tech sector exhibit preferential attachment? What role do governments play in accelerating preferential attachment in the global alliance network in the high-technology sector? The broader purpose of this research is to build theory regarding the role of government in accelerating cooperative inter-organizational activity in strategic industries.

[1] Under preferential attachment, actors have a greater tendency to link with more connected nodes. Thus, as the network evolves a few prominent nodes will emerge as hubs who maintain many ties, while other actors will have few links. This process produces centralized networks.

This article makes two broad contributions. First, it enhances our understanding of the governance role of public and non-profit organizations embedded within private sector networks. Second, it demonstrates the value of integrating complexity science with network governance to reveal how government strategy can capitalize on and promote self-organizing properties of networks. This suggests broader implications for this research in a variety of network governance settings. If decision makers can identify particular network structures best suited to address complex problems, e.g. centralized networks tend to enhance service delivery (Provan and Milward 1995; Milward and Provan 2000), then they might also act with minimal interference to accelerate self-organization around emerging hubs of cooperative activity in broader inter-organizational networks.

## 2. Background

Today the global high-technology sector is structured as a global network of cooperative interactions between organizations. The sector did not always resemble a network but was once dominated by organizations in relative isolation operating as vertical silos. In the semiconductor industry, which is the backbone of the computing and microelectronics industries, a unique innovation trajectory took hold in the latter half of the 20$^{th}$ century, referred to as Moore's Law. Named after former president of Intel, Gordon Moore suggested that the number of transistors packed onto an integrated circuit would double roughly every two years (Epicoco 2011). This non-linear innovation trajectory reached a point during the 1970s and 1980s where firms increasingly began to cooperate on research and development (R&D) in order to remain competitive (Hagedoorn 2001). As the technological landscape shifted toward cooperative behavior between organizations, the Japanese conglomerate system of networked organization, called *Keiretsu*, rapidly achieved market dominance (Lincoln, Gerlach, and Takahashi 1992).

The rising influence of Japanese competitors in the semiconductor industry raised concerns within US business and political circles. Industry leaders and U.S. lawmakers were extremely worried about the loss of majority global market share, which occurred in the mid-1980s, in a strategic industry that was a U.S. invention and dominated by U.S. firms since its inception. As the National Research Council stated in 1992, the semiconductor industry is a *strategic industry* because it is "essential to the nation's well being" (as quoted in Irwin and Klenow, 1994:1201). In this case, the economic and military implications of semiconductor technology are salient to the national interest.

In order to counteract the threat from Japan, the U.S. congress and industry leaders crafted the government sponsored non-profit research consortium, Sematech, abbreviated from Semiconductor Manufacturing Technology. Through the Department of Defense (DOD) and the Defense Advanced Research Projects Agency (DARPA), the consortium received matching funds of $100 million per year for a ten-year period between 1987 and 1996 (Browning and Shetler 2000). The consortium began with participation by fourteen U.S. firms, which accounted for roughly 80% of total U.S. manufacturing capacity, and excluded foreign participants (Browning, Beyer, and Shetler 1995). Members to the consortium contributed one percent of sales, or one million dollars per year, as well as human capital to the fabrication facility in Austin, Texas called Fab One (Browning and Shetler 2000). The primary objectives of the

consortium were to improve manufacturing process quality and to innovate along the miniaturization trajectory of Moore's Law. Leadership and governance were established by the members, and DOD played a very hands-off role (Beyer and Browning 1999). In 1996 Sematech opened membership to foreign firms and DOD matching funds ended, but the consortium continued to function as a non-profit organization.

## 3. Literature Review

This article relies on three broad sources of literature to motivate hypotheses regarding potential effects of government action on inter-organizational networks. The first section applies network governance theory to conceptualize cross-sector inter-organizational cooperation in the high-technology sector and recast the role of Sematech as a network administrative organization (NAO). The second section applies the literature on complexity science, which specifies a unique relational ontology well-suited to the study of networks and highlights specific mechanisms of action. This section describes how preferential attachment provides a unique mechanism for policy makers seeking to enhance self-organization in policy relevant networks. The third section applies the concept of network interventions to suggest why, when, and with what consequences decision makers might choose to intervene in social and organizational networks. Combining these sources of literature suggests how governments might catalyze network emergence. Since the approach is to synthesize somewhat disparate literatures, Table 1 briefly describes each theory, applies the theoretical logic of each to the case, and produces an integrated proposition to motivate the study.

**Table 1. Integrated Theoretical Framework**

<table>
<tr><th>Theory</th><th>Description</th><th>Application to Case</th><th>Integrated Proposition</th></tr>
<tr><td>Network Governance</td><td>Complex public problem resolution requires cooperation between organizations and often across sectors. Network Administrative Organizations (NAO) increase cooperative capacity and enhance network effectiveness in evolving inter-organizational networks.</td><td>The government sponsored non-profit R&D consortium, Sematech, resembles a NAO. The consortium is not itself an actor but coordinates interactions between members and has administrative capacity with respect to a sub-network embedded within a broader global network of alliances.</td><td rowspan="3">Government intervention, via NAO-based link addition, can catalyze cooperative behavior around sets of target organizations in policy relevant inter-organizational networks, accelerating the complex self-organizing process of preferential attachment. Such interventions may be useful for addressing public problems, preventing market failures, and otherwise achieving public ends.</td></tr>
<tr><td>Complexity Theory</td><td>Inter-organizational networks are complex adaptive systems, where catalytic task spaces emerge on fitness landscapes. In such spaces organizational sets combine resources at an accelerated rate. Networks exhibit properties of self-organization, such as preferential attachment in which organizations tend to prefer well-connected partners.</td><td>The global high-technology R&D strategic alliance network is a complex system, where organizations cooperate to combine resources for competitive advantage on a technological fitness landscape. Sematech represents a catalytic task space for resource combination. Organizations tend to seek access to the technological resources of other well-connected organizations.</td></tr>
<tr><td>Network Intervention</td><td>Governments can intervene to affect the development of social networks, adding or deleting nodes, adding or deleting links, and rewiring existing links. Such interventions can accelerate, attenuate, or otherwise modify network development for public ends.</td><td>Sematech represents a type of link addition intervention, where government support for the consortium established new network ties between existing firms in an incipient inter-organizational network.</td></tr>
</table>

*Network Governance*: As government has become less hierarchical and less centralized over the last few decades, the study of cross-sector interorganizational networks has become far more common (Koliba et al. 2019). The New Public Management movement and the emergence of the “hollow state” (Milward and Provan 2000) described a situation where governments increasingly rely on market-based mechanisms to achieve public ends. For example, contracting for products and services (Brown, Potoski, and Van Slyke 2018) and public-private partnerships (Hart 2003; Hodge and Greve 2007) provide viable alternatives to traditional hierarchical within-organization service delivery, i.e. the “make or buy” decision (Brown and Potoski 2003;

Johansson 2015). Similarly, policy networks and collaborative governance regimes emerged to tackle wicked public problems in a variety of areas from natural resource management (Berardo and Scholz, 2010) and environmental protection (Emerson and Nabatchi 2015), to emergency management (Kapucu 2006), and public health (Provan and Milward 1995).

The common thread among these alternative modes of public problem resolution is that they involve cooperation between organizations rather than vertical integration within a single bureaucracy. Sometimes these cooperative inter-organizational linkages are merely a series of dyads with little to no cross-connection. This might be the case with bilateral contracts and public-private partnerships. However, as the number of cooperative linkages between organizations increase, there is a greater probability that a complex network structure emerges. Jones, Hesterly, and Borgatti (1997) suggest that the transaction costs of cooperation between organizations increase their structural embeddedness, which in turn produces the need for governance strategies to adapt, coordinate, and safeguard an evolving network of exchanges. In these scenarios, traditional government approaches to address market failures and support innovation may prove ineffective or counterproductive.

Klijn and Koopenjan (2015:11) define *governance networks* as “more or less stable patterns of social relations between mutually dependent actors, which cluster around policy problems, a policy programme, and/or a set of resources and which emerge, are sustained, and are changed through a series of interactions”. *Governance*, as an activity of government agencies within such networks, is defined as a “set of conscious steering attempts or strategies of actors within governance networks aimed at influencing interaction processes and/or the characteristics of these networks”. Further, the more specific activity of *management*, is defined as “all the deliberate strategies aimed at facilitating and guiding the interactions and/or changing the features of the network with the intent to further the collaboration within the network processes". To the extent that more difficult social problems require more complex organizational responses, these definitions suggest that public problems tend to generate complex patterns of social interaction which might be enhanced or mitigated to achieve broader public ends.

Network governance emphasizes the structural properties of social networks, including constructs like trust, reciprocity, status, prestige, and broader cultural values (Powell 2003). Network governance suggests complex public problems cannot be addressed by individual organizations or government agencies alone, shifting attention to the structural properties of networks rather than management or influence of incentives at the actor level (Agranoff 2006). As such network governance often crosses the boundaries of public, private, and non-profit sectors. However, organizations party to governance networks are neither strongly managed by government agencies, nor are they fully adversarial in the market sense of competitive advantage (Ansell and Gash 2008). Rather, actors remain autonomous but are non-trivially dependent on access to the heterogeneously distributed resources of their network partners.

As Provan and Kenis (2008) suggest, network governance has tended to display three generic modalities: the participant governed model, the lead organization model, and the network administrative organization (NAO) model. The first is a decentralized model having no lead organization, while the second is a centralized network model where a dominant organization leads a network of inter-organizational relationships. In contrast to both, the NAO represents a middle ground scenario, where a new organization is established to govern the network of relationships, but which is external in some sense to the industry it is designed to govern. In the private sector context, the NAO is often a non-profit organization specifically designed to coordinate activities between private sector organizations (Human and Provan 2000). As a non-profit, the NAO is not a direct competitor with members of the broader inter-organizational network. The NAO may be conceptualized as a type of organizational intermediary or "collaborative platform" designed to facilitate cooperative behavior between organizations (Ansell and Gash 2018).

Wardenaar, de Jong, and Hessels (2014) suggest the network governance approach provides a useful conceptualization of strategic research consortia in science and technology policy. In this article, we conceptualize the government sponsored research consortium, Sematech, as a mode of network governance closely related to the NAO model, where Sematech has administrative capacity with respect to a sub-network within a broader global network of alliances (see section two for details about Sematech). This conceptualization suggests that government intervention may provide the capacity necessary to enhance the effectiveness of networks of organizations in rapidly changing technological landscapes. By focusing on the structural properties of social networks, the network governance approach may represent a more sustainable long-term strategy for achieving public ends in science and technology networks.

*Complex Systems:* The reasons that policy makers should be attentive to the properties and dynamics of whole networks, conceptualized as complex systems, may not be immediately apparent. As Provan, Fish, and Sydow (2007: 480) suggest, "Only by examining the whole network can we understand such issues as how networks evolve, how they are governed, and, ultimately, how collective outcomes might be generated". However, insufficient attention has been drawn to whole networks in public administration or to recasting whole network properties and dynamics as complex systems, *per se*. This section suggests conceptualizing policy relevant inter-organizational networks in terms of complexity theory may provide policy makers with previously unexplored concepts that benefit the implementation of public policies and programs and lead to more effective outcomes given public objectives.

Complexity theory suggests attention to unique generative mechanisms that manifest at the systems level. Thurner, Klimek, and Hanel (2018:22) define complex systems generically across natural and social phenomena as “co-evolving multilayer networks”. This definition of complexity suggest that complex systems are understood specifically as overlapping and interacting sets of dynamic networks. Ladyman, Lambert, and Wiesner (2013) point to the concepts of numerosity and interaction, where numerosity refers to the fact that complex systems involve numerous elements, actors, or agents, while interaction simply means that these elements are all interacting with each other in a non-trivial way. Miller and Page (2009:44) point to the concepts of self-organization and emergence, where “localized behavior aggregates into global behavior that is in some sense disconnected from its origins”. Self-organization is compatible with decentralization in network governance. In contrast to top down administration by central agencies, the local partnering behavior between organizations gives rise to a less centralized and more adaptive structure of interaction where network structure exhibits properties of self-organization such as reciprocity, closure, and homophily (Robins, Lewis, and Wang 2012). Complex interaction can further lead to the emergence of mechanisms such as positive and negative feedback loops, which can enhance or diminish behavior among local actors (Miller and Page 2009). Attention to feedback is also important to the anticipation of the potential consequences of public action in complex arenas.

Kauffman (1993) suggested that complex systems can be characterized as fitness landscapes. In such systems, organizations cooperate to summit peaks of fitness on technological landscapes, where recombination of technology produces novel innovations. A feature of evolution on such landscapes is called “catalytic task spaces”. While the original concept was designed to describe chemical interactions that give rise to biological processes, the notion can be generalized to technology landscapes, where organizations combine their respective resources and recombine existing technologies to produce new innovations. The catalytic task space may be thought of as a local cluster within the landscape where recombination occurs at an accelerated rate. Kauffman’s (1993) notion of a catalytic task space fits well with the idea that government might act as a network catalyst. The Oxford English Dictionary defines a catalyst as “A substance which when present in small amounts increases the rate of a chemical reaction or process but which is chemically unchanged by the reaction; a catalytic agent”. In this sense, if governments create local clusters of cooperative behavior, such clusters may *catalyze* a broader process of cooperation on the landscape. While, the metaphor breaks down over whether the government remains unchanged in the process, the rate of increase in a chemical reaction maps on well to the notion of increased cooperative behavior in, for example, government sponsored consortia.

Complexity has been applied to numerous disciplines outside of the natural sciences. For example, Arthur (1996) applies complexity to the study of economics through increasing returns to technological advancement. Complexity concepts have also been applied by public policy and administration scholars, including nonlinearity (Morçöl 2012; McGee and Jones 2018), self-similarity, feedback (Eppel 2017), self-organization (Comfort 1994; Berardo and Scholz 2010), fitness landscapes (Rhodes and Dowling 2018), and preferential attachment (Carboni and Milward 2012). However, there has been insufficient attention to the nature of complex systems and the application of statistical modeling of social networks in the public policy and administration literature (Robins, Lewis and Wang 2012). Observations regarding decentralization of governance and increasing cross-sector cooperation abound, yet extant concepts from complexity theory remain underutilized given their potential for describing these very situations. This article applies a set of complexity concepts to understand network governance generically, as well as specific mechanisms of complex systems to motivate hypotheses regarding government intervention in inter-organizational networks. Further, this study builds on Browning and Shetler's (2000) qualitative analysis which specifically characterized Sematech in terms of complexity theory and self-organization.

The specific mechanism of interest in this research is preferential attachment. Preferential attachment describes a process where network actors tend to seek out connections with already well-connected actors (Barabasi and Albert, 1999; Newman, 2001) and where popularity tends to generate increasing popularity (Robbins, Lewis, and Wang 2012). The concept has many permutations and precursors such as the *Mathew Effect* described by the sociologist Robert K. Merton (1968), i.e. "the rich get richer, and the poor get poorer". Arthur (1986) describes a similar process, suggesting that technological advancement tends to generate positive feedback producing increasing returns to investment. Preferential attachment is a distinct process from the Mathew Effect and increasing returns, however, because it focuses on social connections. This process leads to a positively skewed distribution of network connections. Such networks often follow a power law distribution and are characterized as scale-free when 1) preferential attachment is present, and 2) the network is continuously expanding (Barabasi and Albert 1999). However, the question of whether many social networks are actually scale-free has come under criticism (e.g. Broido and Clauset 2019) and is beyond the purpose of this article. We apply the concept of preferential attachment to suggest that organizations within inter-organizational networks tend to seek and form alliances with already well-connected organizations.

Although preferential attachment is a prominent concept in network analysis, its application to the field of public policy and administration has been limited. Literature has discussed preferential attachment as a structural feature of social networks (Robins, Lewis and Wang 2012; Weare, Lichterman and Esparza 2014). Sun and Cao (2018) used preferential attachment to understand the structure and function of Chinese research and development (R&D) policy networks, finding that government agencies were more powerful as a result of preferential attachment. They argued that policymakers could capitalize on preferential attachment to strengthen interagency collaborations. Similar studies on R&D policy networks found that repeat participation within the network (Protogerous, Caloghirou, and Siokas 2010) and a broader process of self-organization (Biggiero and Angelini 2015) were factors leading to preferential attachment within policy networks. Lake and Wong (2009) suggested that preferential attachment could enhance political power but also pose challenges when powerful actors are removed from centralized networks. Concerns have also been raised regarding network stability (Carboni and Milward 2012) and hindrances to the flow of information (Lyles 2015). Schilling and Fang (2014) proposed that “moderately hubby” network forms prevent instability, while Koliba et al. (2017) suggest improving administrative support in order to build trust between less centralized actors.

R&D alliance networks in the high-technology sector are like other social networks in many respects, featuring common processes such as reciprocity, homophily, and preferential attachment. Preferential attachment may occur for functional or institutional reasons. Cooperative extensions of the resource-based view of the firm (RBV) suggest that firms seek out partnerships in order to gain access to heterogeneously distributed resources which might be applied for joint advantage on competitive landscapes (Eisenhardt and Shoonhoven 1996; Das and Teng 2000; Scott and Thomas 2017). Thus, firms will tend to seek out partnerships with well-connected firms in order to gain access to their deep resource portfolios. The benefits of network partnership can extend well beyond resource acquisition, allowing organizations to shift blame and at times conceal operations (Jensen 2016). Further, Fligstein and McAdam (2011) suggest firms are attracted to partnerships with other firms, and/or groups of firms, that possess the social resources to mobilize and influence action around a collective purpose. Preferential attachment highlights the relevance of social resources and social capital (Lin 1999), as well as drivers of alliance formation related to cognitive and institutional processes (Inkpen and Tsang 2005; Powell et al. 2005). For example, Stuart (1998) showed that prestige is a factor driving alliance formation in the semiconductor industry. For these reasons, we hypothesize that, in the semiconductor industry, organizations with many alliances will tend to form even more alliances, i.e., a preferential attachment process drives network evolution.

> *H1: The semiconductor industry network exhibits preferential attachment in the distribution of strategic alliances.*

*Network Interventions*: Finally, a small but emerging literature on network interventions is relevant to the current study. Valente (2012:49) defines network interventions as "the process of using social network data to accelerate behavior change or improve organizational performance". Building on network theory developed in the health sector, Valente suggests that governments might intervene to stimulate the emergence of a network, to alter the structure of an existing network, or to even break up an existing network. As Valente states,

> "When network data indicate that the network is nonexistent, too fragmented, too centralized, or otherwise dysfunctional, there is a need for network change. The interventionist should use induction or alteration techniques to create a network amenable to change. Once the network is built or restructured, identification and segmentation tactics can be used to accelerate change." Valente (2012,52)

The basic propositions that Valente advance is that understanding network structure is important and governments might take specific modalities of intervention in order to achieve public ends. These include, among others, adding or deleting nodes, adding or deleting links, and rewiring existing links. One of the most common types of intervention strategies involve identifying key players in order to elevate them to leadership positions for policy implementation (Valente et al. 2015). We take a different approach, suggesting that government action, *via* Sematech, represents a type of link addition intervention, where the consortium establishes network ties between existing firms in an incipient network.

Scott (2016) suggests that these types of collaborative network interventions have not been systematically examined in policy studies. Scott shows that participation in government sponsored collaborative groups enhances further collaborative ties between organizations by reducing the transaction costs of cooperation in the network. We suggest a similar logic applies to Sematech as a mode of network intervention and governance, where the consortium lowered cooperation costs that enabled large scale resource combination.

We combine insights from governance networks, complexity theory, and network intervention to advance the hypothesis that governments might catalyze cooperative behavior in policy relevant sectors by accelerating the process of preferential attachment around a set of policy relevant target firms in an emerging network. By creating a task space around a set of target organizations, governments might catalyze a broader process of preferential attachment. In the present case, members of the government sponsored research consortium, Sematech, are the policy relevant targets of intervention.

> *H2: Government intervention through NAO-based link addition accelerates preferential attachment around a set of public policy relevant target firms.*

Since the implementation of the Sematech program occurred over time, the effects are likely time dependent. Following existing literature on Sematech (Browning and Shetler 2000; Carayannis and Alexander 2004), we distinguish primarily between an implementation phase, a maturity phase, and a post-DOD phase of implementation. Government sponsorship likely enhances the network capacity of the consortium, while the sponsorship and exclusion of foreign firms from the consortium likely enhances the prestige of the members. Such increases in social capital further contribute to a process of preferential attachment. Thus, we hypothesize that the preferential attachment effect will be strongest during the implementation and maturity phase of government sponsorship.

> *H3: The strongest Preferential attachment effect of government intervention occurs during the implementation and maturity periods of government sponsorship; a weaker effect is expected in the post-sponsorship period.*

## 4. Data, Methods, and Variables

The research design of this study is a time series analysis of a single case, the semiconductor industry from 1987 to 1999, using longitudinal data on the strategic alliances and attributes of 451 firms in the global semiconductor industry. Government intervention is operationalized by participation of organizations in the DOD-sponsored non-profit R&D consortium, Sematech. This article employs two types of stochastic network analysis methods appropriate to analyzing two distinct types of network data. Further, we add a comparison of these methods and their modeling terms that may be of methodological interest to the future work of network analysts in public administration.

*Data Sources:* Data on network connections between firms were gathered from two principal sources. Data from 1987-1989 were gathered from public announcements compiled through press releases and other public news announcements using the SIC code 3674 for semiconductors and related devices in the Lexis-Nexis database. Alliance data in the strategic management literature are commonly measured through public announcements (Schilling 2009). One of the limitations of announcement data is the lack of information regarding ongoing alliances or dissolution of alliances. Beginning in 1990, a new data source became available. Data from 1990 to 1999 were gathered from Integrated Circuit Engineering Corporation's (ICE) data books, *Profiles: A Worldwide Survey of IC Manufacturers and Suppliers* (ICE 1991-2000).[2] These reports include strategic profiles for firms each year and include information on ongoing alliances, allowing for the tracking of tie formation and dissolution. Network connections

---

[2] ICE was a well-regarded market research firm focused specifically on the analysis of the semiconductor industry. This research is now conducted by a firm called IC Insights, which produces the same company profile data currently called *Strategic Reviews* (data may be obtained from www.icinsights.com). These data have also been previously used in the strategic management literature (Leiblein, Reuer, and Dalsace 2002; Leiblein and Miller 2003; Leiblein and Madsen 2009)

between firms are operationalized by each focal firm's list of partners and their business relationships, including co-development agreements, cross-licensing agreements, equity investments, joint ventures, marketing agreements, and mergers. In the business literature, these types of relationships are characterized as strategic alliances (Eisenhardt and Schoonhoven 1996). The alliances represent reciprocal exchange of resources, including financial resources, social and human capital, technology, and information. The data also included firm level data on sales and headquarters location. Sematech members were identified by contacting Sematech (see online appendix for a list of members). The strategic alliance data were used to create undirected binary networks for each year between 1987 to 1999. These time periods were chosen because the implementation of Sematech started in 1987, the maturity of Sematech began roughly in 1991, and the post-DOD sponsorship era of Sematech occurred with DOD-exit in 1996 (Browning and Shetler 2000).

*Stochastic Network Analysis Methods:* We use two different stochastic network analysis methods, which we attempt to model in parallel form: 1) exponential random graph modeling (ERGM) of a network of 283 organizations; and 2) stochastic actor-oriented modeling using RSiena (SIENA) of a network of 451 organizations. The difference in sample size for the two methods is due to the size of the network at different points in time, where the ERGM covers the time period from 1987-1990 and the SIENA models covers 1991-1999.[3] Ideally, we would model all of the years together in a single SIENA model. However, as noted above, the early alliance data gathered between 1987 and 1989 provides information on tie formation only and not ongoing alliances. This prevents us from being able to track the evolution (formation and dissolution) of ties during these years. Thus, for the early period we used an ERGM on a pooled network that consisted of all ties formed between 1987 and 1990. We chose to include 1990 in the ERGM and start the SIENA models in 1991 so that each SIENA model is a symmetrical 3-year time slice where the mid-point between the four models is the year of DOD-exit in 1996. Further, Sematech membership is static from 1987-1990, and Sematech experienced a major shift in consortium strategy in 1991 which represents the shift into the maturity phase (Browning and Shetler 2000). Conversely, data from 1990 forward included ongoing alliance information, which allowed for the use of SIENA to model the network dynamics over time. In contrast to the ERGM, these data included 451 firms who operated in the semiconductor industry at any point during that time period. Firm exit and entry into the industry is controlled for in the model through the use of structural zeros (Ripley et al. 2019). We include analogous modeling terms in both approaches to maintain continuity between the two different methods. As a robustness check, we include the pooled 1987-90 data as the first year in the SIENA model in the online appendix.

---

[3] The sample size for the SIENA models is a conservative undercount of the actual number of nodes and edges. This is because the SIENA model requires a single list of nodes for the entire period. The Siena models are restricted to those firms listed in the ICE *Profiles* data books. In the ERGMs we were free to include nodes that were partnered with firms covered in the organizational data, but which were not themselves featured with a profile.

ERGM is a method for estimating the probability of tie formation between nodes in a network (Lusher, Koskinen, and Robins 2013). Ties that form within a network are not independent of one another creating dependencies that are not appropriately accounted for with traditional statistical methods, such as logistic regression (Butts 2008, Snijders 2011). ERGMs approximate a maximum likelihood estimate of the coefficients through a Markov Chain Monte Carlo (MCMC) simulation process and take the following form (Robins 2007):

$$\Pr(Y = y) = \left(\frac{1}{k}\right) exp\left\{\sum_{A} \eta Ag\ A(y)\right\}$$

Multiple configuration types, including structural effects, nodal attributes, and homophily effects can be contained in gA(y). The structural effects are endogenous terms whose value depends on the configuration of the network. Unlike actor attribute effects and homophily effects, which only rely on the two members of the dyad to determine their value, structural effects are dependent on the rest of the ties in the network. These terms are included to capture self-organizing properties of the network. The model parameters, $\eta A$, estimate the relative importance of each configuration. The parameters are estimated and updated until the observed networks become central in the distribution of networks simulated from the current model. In other words, the model reaches convergence when the model parameters make the observed network the most probable (Lusher, Koskinen and Robins 2013).

The data from 1990-1999 include information on ongoing alliances and alliance dissolution, permitting stochastic actor-oriented modeling. In order to explore the formation and evolution of cooperative relationships in the semiconductor industry, we use the RSiena (Simulation Investigation for Empirical Network Analyses) program in R, simply referred to as SIENA models. SIENA models are an appropriate analytic tool when one has a panel of network observations. SIENA models are continuous-time Markov chain models where tie changes are determined by the current state of the network and whose parameters are estimated through a series of simulations (Snijders et al. 2010). The models are actor-oriented in the sense that the network evolves through each actor's decisions about which ties to dissolve, maintain, and form in the network. While data consists of snapshots of the semiconductor industry, changes in the network take place continuously during the time elapsed between consecutive periods.

SIENA models have two main components. A timing process which determines the number of opportunities an actor has to update the network (i.e., create, maintain, or dissolve a tie) and a choice process driven by an objective function which determines the probability that an actor makes a particular update. Thus, SIENA models estimate the underlying and unobserved network evolution through a series of micro-steps taken by each actor in the network. Based on the language and notation in Ripley et al. (2019, p. 119), the network objective function for actor $i$ is defined as:

$$f_i^{net}(x) = \sum_k \beta_k^{net} \, s_{ik}^{net}(x)$$

where $\beta_k^{net}$ are the parameters and $s_{ik}^{net}$are the effects. Based on the network objective function, each actor's utility of dropping an existing tie, forming a new tie, or maintaining his or her existing network is assessed, and a probability assigned. If the value of the objective function increases for the formation of a particular tie, then the probability of that action occurring also increases.

SIENA models were originally developed for directed networks based on the assumption that a tie's existence is determined by the sender of the tie. Recent developments have extended these models to non-directed networks, such as those found in the semiconductor industry as well as in a wide variety of governance networks. As discussed by Snijders and Pickup (2016), non-directed networks require adjustments to the timing process and the choice process.[4] We rely on a one-sided opportunity process and a mutual choice process to model the dynamics of the semiconductor industry. We selected this combination because a one-sided initiative with mutual confirmation "is in most cases the most appealing simple representation of the coordination required to create and maintain non-directed ties" (Snijders and Pickup 2016, p. 233). We believe this methodological choice best approximates the process of strategic alliance formation in practice, e.g. co-development agreements as a mutual contract representing a mutual pooling of resources but for strategic objectives (e.g. Eisenhardt and Shoonhoven 1996). Firm entry and exit (i.e., composition change) in the semiconductor industry is controlled for in the model through structural zeros (see Ripley et al. 2019).

[4] These new options arise because the formation of a tie can longer be based solely on the utility considerations of the sender of the tie. In non-directed networks, senders and receivers cannot be differentiated and thus one must decide how to take into account the utility of each actor. According to Snijders and Pickup (2016) the timing process for non-directed networks has two possibilities: one-sided initiative (where a single actor is randomly chosen); or two-sided opportunity (where an ordered pair of actors is randomly selected). More importantly, the choice process has three options: dictatorial (where one actor's utility determines the tie); mutual (where both actors must agree on the tie; in the sense that it is beneficial to both); and compensatory (where a combined objective function is established for each pair of actors).

*Model Terms and Variables:* In order to establish continuity across the two modelling techniques, we employ parallel modelling strategies with analogous terms between the two approaches. Both the ERGM and SIENA models employ three types of effects: structural effects, actor attribute effects, and homophily effects. We provide generic variable names for analogous model terms across ERGM and SIENA models described in Table 2 as well as a brief description of their function.

For structural effects in the ERGMs, the variable of interest for our first hypothesis is the geometrically weighted degree distribution, *gwdegree*. This term captures the skewness of the distribution, where a negative estimate indicates a highly skewed distribution (Levy 2016, Hunter 2007). This term serves as a measure for a general preferential attachment process in the early period of the alliance network. The *gwdegree* term, as parametrized, captures an anti-preferential attachment process and thus a negative coefficient provides support for H1 (Hunter 2007). For the structural effects in our SIENA model we rely on a popularity effect, called *inPop*, to capture the tendency for actors to form ties with others who already have many ties. The popularity effect operationalizes the process of preferential attachment hypothesized in H1. Unlike the *gwdegree* term, a positive coefficient on *inPop* indicates preferential attachment.

For node attribute effects in the ERGM we use the *nodefactor* term; and, in the SIENA model we use the *egoPlusAltX* term for undirected networks. These both permit the estimation of categorical variables on the probability of tie formation. The key variable of interest is whether the node is a Sematech member during time *t*. We take this term as our basic proxy for Sematech induced preferential attachment, as it models the tendency for certain organization types to be nominated more than others (i.e., is there preferential attachment to certain node types). These Sematech terms serve as our proxy measures for the test of H2. In the ERGM model there is no change in Sematech membership for the pooled networks between 1987 and1990. For the SIENA models Sematech membership changes from year to year.

Control variables include main effects for country headquarters, specifically focusing on USA and Japanese (JPN) headquartered firms, as well as homophily effects of country and Sematech membership. We chose to include variables for the USA and JPN because those are the two dominant nations during the period, and more specifically, to highlight that USA public policy was specifically directed at JPN competition. Firms headquartered in these two countries account for roughly 80% of global semiconductor sales during the time period (SIA 2016). We also control for other self-organizing processes prevalent in networks, such as transitivity (i.e. the tendency for actors to form subgroups or triads where all three nodes are connected). In the ERGM this is captured by geometrically weighted edgewise shared partner distribution term, called *gwesp*; and, in the SIENA model this is captured by the *transTriad* term. Lastly, a density parameter is also included in the models. In the ERGM this is the *edges* term; while the SIENA model is *degree(density)*. This variable functions similarly to an intercept term in a standard linear model and captures the overall tendency for ties to form in the network.

## Table 2 – Description of Variables

| Independent Variables | Definition | ERGM Term | SIENA Term |
|---|---|---|---|
| Pref.Attach | Orgs with ties gain more ties | gwdegree | inPop |
| Sematech Effect | Sematech members gain more ties | nodefactor | egoPlusAltX |
| **Control Variables** | | | |
| Sematech Homophily | Sematech members form ties with Sematech members | nodematch | sameX |
| USA Effect | USA orgs form more ties | nodefactor | egoPlusAltX |
| USA Homophily | USA orgs form ties with USA orgs | nodematch | sameX |
| JPN Effect | JPN orgs form more ties | nodefactor | egoPlusAltX |
| JPN Homophily | JPN orgs form ties with JPN orgs | nodematch | sameX |
| Org. Size | Three year moving average of sales for org-year | -- | egoPlusAltX |
| Transitivity | Tendency of network to form triads | gwesp | transTriad |
| Density | The overall density of the network | edges | degree(density) |

Table Notes: the table shows the name of each variable, the definition, the ERGM term name, and the SIENA term name. ERGM term for Org.Size is blank because the variable was not used for the ERGM in which the nodecov function requires non-missing data for continuous variables.

## 5. Results

The results of the exponential random graph models (ERGM) used in the implementation period (1987-1990) are below in Table 3. The model and goodness-of-fit diagnostics are in the online appendix, showing good convergence statistics. The first model shows the base ERGM with the degree distribution term (Pref.Attachment). The negative estimate indicates high skewness in the distribution (Levy 2016, Hunter 2007), which is our proxy for general preferential attachment in network tie formation. This result provides support for H1 during the early implementation period. In the second model, we add the main effect term (Sematech Effect), which captures the tie formation effect of membership in the government sponsored network administrative organization. The estimate is significant and positive, and Pref.Attachment has now lost its significance. This indicates that the main effect of Sematech accounts for a portion of the skewness in the degree distribution. In other words, to the extent that the network has a right skew, those high degree nodes can be partially accounted for by Sematech membership. The third model adds the transitivity term, control variables for country headquarters, and homophily terms for Sematech members and firms of the same country headquarters. The effect of Sematech remains positive and significant, while Pref.Attachment has reversed sign and is now significant. These results provide support for H2 during the implementation period.

**Table 3 – ERGM Models, Implementation Period (1987 to 1990)**

| | Model 1 | | Model 2 | | Model 3 | |
|---|---|---|---|---|---|---|
| Density | -3.998*** | [0.000] | -5.019*** | [0.000] | -7.529*** | [0.000] |
| | (0.069) | | (0.122) | | (0.299) | |
| Pref.Attachment | -1.463*** | [0.000] | -0.178 | [0.388] | 1.345*** | [0.000] |
| | (0.158) | | (0.207) | | (0.268) | |
| Transitivity | | | | | 0.857*** | [0.000] |
| | | | | | (0.082) | |
| USA Effect | | | | | 1.921*** | [0.000] |
| | | | | | (0.190) | |
| USA Homophily | | | | | -1.874*** | [0.000] |
| | | | | | (0.267) | |
| JPN Effect | | | | | 0.787*** | [0.000] |
| | | | | | (0.184) | |
| JPN Homophily | | | | | 0.179 | [0.320] |
| | | | | | (0.180) | |
| Sematech Effect | | | 2.177*** | [0.000] | 1.264*** | [0.000] |
| | | | (0.099) | | (0.145) | |
| Sematech Homophily | | | | | 0.222 | [0.168] |
| | | | | | (0.161) | |
| AIC | 4691 | | 4224 | | 3982 | |
| BIC | 4708 | | 4250 | | 4059 | |

Table Notes: *** p < 0.001, ** p < 0.01, * p < 0.05; exact p-values in brackets; 0.000 is used for p-values below 0.001; standard errors in parentheses. The network includes 283 organizations. The models are for the implementation period, with pooled yearly networks from 1987-1990. Model 1 shows the ERGM only with the general preferential attachment measure(gwdeg). Model 2 adds the Sematech main effect. Model 3 includes all controls. Goodness of fit diagnostics are shown in the online appendix. Models used a seed of 0 in R code for replicability; ERGM models tend to produce slightly different results on each run.

Interestingly, during this period, USA firms appear to have higher propensity for alliance formation than JPN headquartered firms; yet USA firms had a negative propensity for forming ties with each other. This is consistent with the extremely competitive nature of USA firms, also indicating the USA firms have strong alliance formation tendencies with foreign firms despite the essentially protectionist nature of Sematech. Finally, Sematech members appear to have a positive tendency to form ties with each other during this period. However, the estimate is only marginally significant.

Next, we analyzed four distinct time segments, within the maturity period and the post-DOD period of the evolution of the collaboration network, using stochastic actor-oriented models in RSiena (SIENA). Table 4 shows four SIENA models for each three-year time period. Each time period is modeled separately to examine how the relevance of the key variables may change during the different periods: Model 1, 91-93; Model 2, 93-95; Model 3, 95-97; and Model 4, 97-99. All four models indicated good convergence statistics. The absolute values of the convergence t-ratios were all less than 0.1 and the overall convergence ratio was less than 0.25 (Snijders, van de Bunt, and Steglich 2010). Goodness-of-fit diagnostics for the SIENA models are in the online appendix. In support of HI, all four models show a positive and significant general effect on Pref.Attachment. However, our proxy for Sematech induced preferential attachment (Sematech Effect) only showed a positive significant effect in Model 1 providing partial support for H2 during the maturity period.[5] In Model 2 the estimate reverses but is insignificant. Then, in Model 3 the sign remains negative but becomes significant. The sign becomes positive in Model 4 but is insignificant. The reversal of sign during the post-DOD period suggest that the attractiveness of Sematech membership on alliance formation were strongest during the implementation period and during the first half of the maturity period which supports H3. The reversal of significance between Model 1 and 2 may be due to several members leaving the Sematech consortium, while the increase in significance of the negative estimate in Model 3 supports H3.

---

[5]As a robustness check, the data from 1987-90 were also included in an initial SIENA model combined as the first year in the model, included in the online appendix. This also shows consistent results to Model 1 in Table 4.

## Table 4 – SIENA Models, Maturity to post-DOD period (1991 to 1999)

| | Maturity Period | | | | Post-DOD Period | | | |
|---|---|---|---|---|---|---|---|---|
| | Model 1 | | Model 2 | | Model 3 | | Model 4 | |
| | Yrs.91-93 | | Yrs.93-95 | | Yrs.95-97 | | Yrs.97-99 | |
| **rate period 1** | 0.821*** | [0.000] | 1.308*** | [0.000] | 0.618*** | [0.000] | 0.823*** | [0.000] |
| | (0.094) | | (0.129) | | (0.071) | | (0.085) | |
| **rate period 2** | 0.933*** | [0.000] | 1.081*** | [0.000] | 1.222*** | [0.000] | 1.180*** | [0.000] |
| | (0.106) | | (0.088) | | (0.112) | | (0.140) | |
| **Density** | -3.314*** | [0.000] | -3.302*** | [0.000] | -3.448*** | [0.000] | -4.733*** | [0.000] |
| | (0.428) | | (0.401) | | (0.411) | | (0.471) | |
| **Transitivity** | 0.330** | [0.006] | 0.614*** | [0.000] | 0.563*** | [0.000] | 0.774*** | [0.000] |
| | (0.121) | | (0.086) | | (0.079) | | (0.089) | |
| **Pref.Attachment** | 0.123*** | [0.000] | 0.107*** | [0.000] | 0.097*** | [0.000] | 0.082*** | [0.000] |
| | (0.019) | | (0.013) | | (0.011) | | (0.016) | |
| **USA Effect** | -0.045 | [0.829] | -0.284* | [0.026] | 0.503** | [0.005] | 0.042 | [0.809] |
| | (0.207) | | (0.128) | | (0.181) | | (0.172) | |
| **USA Homophily** | -0.198 | [0.339] | -0.124 | [0.369] | -0.450* | [0.011] | 0.094 | [0.568] |
| | (0.207) | | (0.138) | | (0.176) | | (0.164) | |
| **JPN Effect** | 0.816* | [0.020] | 0.722** | [0.003] | 1.279*** | [0.000] | 1.048** | [0.001] |
| | (0.349) | | (0.243) | | (0.306) | | (0.321) | |
| **JPN Homophily** | 0.140 | [0.661] | 0.814*** | [0.000] | 1.156*** | [0.000] | 0.669* | [0.028] |
| | (0.319) | | (0.239) | | (0.301) | | (0.304) | |
| **Sematech Effect** | 0.825* | [0.018] | -0.690 | [0.066] | -0.600 | [0.066] | 0.287 | [0.399] |
| | (0.350) | | (0.375) | | (0.327) | | (0.341) | |
| **Sematech Homophily** | 0.403 | [0.236] | -0.417 | [0.232] | -0.847** | [0.006] | 0.139 | [0.673] |
| | (0.341) | | (0.349) | | (0.308) | | (0.328) | |
| **Firm Size Effect** | -0.144 | [0.366] | -0.560*** | [0.000] | 0.074 | [0.255] | 0.015 | [0.803] |
| | (0.160) | | (0.129) | | (0.065) | | (0.059) | |
| **Iterations** | 2867 | | 3060 | | 3060 | | 3060 | |

Table Notes: ***p<0.001,** p<0.01,*p<0.05; exact p-values in brackets; 0.000 is used for p-values below 0.001; standard errors in parentheses. The networks include a total of 451 organizations. The first two models represent the maturity period, the second two models represent the post-DOD period. DOD exit occurred in 1996. Models 2-4 include the last year of the previous model as SIENA conditions parameter estimates on the initial observation. Thus, we asses change in network ties that occurs between the final time point of the previous model and the subsequent observations for the current time period. Sematech Effect represents the target preferential attachment effect. Goodness of fit diagnostics are shown in the online appendix. Models used a seed of 0 in R code for replicability; SIENA models tend to produce slightly different results on each run.

In summary, the results indicate strong support for H1, moderate support for H2, and moderate support for H3. Table 5 summarizes our main results.

**Table 5 – Summary of Empirical Results**

| | Model Years | | | | |
|---|---|---|---|---|---|
| | 1987:1990 | 1991:1993 | 1993:1995 | 1995:1997 | 1997:1999 |
| **Pref. Attachment** | + | + | + | + | + |
| **Sematech Effect** | + | + | ns | - | ns |

Table Notes: the positive coefficient in the ERGM model for 1987:1990 indicates a tendency away from preferential attachment, while the positive coefficients for the SIENA models indicate a tendency toward preferential attachment.

The control variables also showed interesting estimates. The positive and negative signs on Sematech homophily appear to correlate with the main Sematech Effect, suggesting that when Sematech members form alliances with each other, they also seek alliances with others. Sematech homophily is strongly negative in the immediate post-DOD period, perhaps suggesting that Sematech substitutes for alliance formation between USA firms and foreign firms. JPN firms show a higher propensity for collaboration across the time periods and a higher propensity to collaborate with other JPN firms, while USA firms only appear to show positive tie formation in the immediate post-DOD period and show a consistent negative propensity to collaborate with other USA firms.

## 6. Discussion

The study poses the theoretical proposition that government intervention, *via* NAO-based link addition, can catalyze cooperative behavior around target organizations in policy relevant inter-organizational networks, accelerating the complex self-organizing process of preferential attachment. This proposition entails three hypotheses regarding alliance formation in the semiconductor strategic alliance network: H1) a general preferential attachment process is present in the distribution of network ties; H2) a government sponsored consortium, structured as a network administrative organization (NAO), accelerates this process; and H3) the effects are time dependent and differ during and after government sponsorship. The results provide support for the proposition and hypotheses, showing strong support for a generic process of preferential attachment in the network, suggesting that government intervention helped to build the network's capacity around target organizations in the early period, and also suggesting that as DOD stepped back from the network an ongoing process of self-organization contributed to the development of the network.

More specifically, the analysis suggests that preferential attachment operates in the industry network generically, but the situation regarding target organizations is more nuanced. The results indicate that NAO membership is associated with rapid tie formation in the implementation and early maturity periods of the development of the program. While, the Sematech effect appears to diminish and even reverse as the consortium proceeds through the maturity period and then enters the post-DOD period, the general preferential attachment effect remains. Thus, US firms who were members of Sematech benefited from additional tie formation and collaboration in the early years, and while they cemented into positions of popularity a general preferential attachment process continued to increase the centralization and structural cohesion around this core group of firms. This raises the point that over time DOD could step back and the network could still function. However, a related explanation might suggest that the reversal of the consortium effect in the post-DOD period may be due to saturation in alliance formation as target firms reach a threshold of utility for new alliances, which supports observations regarding diminishing returns to alliance portfolio size (Oxley 1997; Lahiri and Narayanan 2013). As such, there may be limits to government sponsorship efforts aimed at firms with large alliance portfolios. In either case, the study underscores the timeliness with which network interventions must take place. Those aimed at the early stages of network formation may be more effective than those aimed at networks with well-established governance routines and safeguarding mechanisms (e.g. Jones, Hesterly, and Borgatti 1997).

This study contributes to the integration of complexity science with the network governance literature in public policy and administration (Comfort 1994; Morçöl, G.K. 2012; Eppel, 2017) by suggesting preferential attachment is a property of governance networks that might be accelerated, attenuated, or generically manipulated to achieve public goals. Scholars have noted that service delivery tends to perform better in centralized networks (e.g. Provan and Milward 1995; Milward and Provan 2000) but have not examined how government can promote the emergence of centralized networks, for example, by capitalizing on the self-organizing mechanism of preferential attachment. As such, scholars and decision makers should take notice of the properties and dynamics of complex systems in order to anticipate the effects of policy implementation through cross-sector inter-organizational networks. As Comfort (1994) recognized, self-organization on fitness landscapes requires a balancing act of structure and flexibility. As she eloquently stated, “The vital but elusive characteristic of self-organization is its spontaneity. While influenced by the actions of other organizations or groups, it cannot be imposed by external regulation nor can it be suppressed by perpetual chaos” (410). In this sense, the task of interested decision makers is to maintain the necessary balance between structure and chaos for cooperative networks to flourish. This is consistent with the relatively hands-off approach of DOD support for the Sematech research consortium. Rather than creating new public bureaucracies, providing direct subsidies to private firms, or enacting trade barriers in the international system, this research suggests that governments might act upon social structure rather than the agents themselves.

Further, this study synthesizes the literature on network interventions (Valente 2012) and complexity with network governance in order to more fully explore the theoretical underpinnings of inter-organizational networks, as well as the role of governments within these networks. We conceptualize the NAO approach in the Sematech case as a tie-addition intervention but further suggest a positive feedback loop where adding ties among existing node sets may accelerate the development of more ties. This is evident because the propensity of USA firms to form ties with other USA firms is largely negative or insignificant across all time periods, suggesting that the NAO permits cooperation between firms which otherwise are highly competitive. A tie addition strategy through alliance sponsorship may provide the needed governance structure to reduce cooperation costs and permit resource combination.

Further, the global nature of the high-technology sector network suggests governance effects may be country and culture specific, supporting or dampening the effects of public programs for inter-organizational collaboration. The results reinforce the general notion that USA firms tend to be highly competitive, and indeed, appear to form cooperative ties with each other with a lower propensity than JPN firms during the maturity and post-DOD periods. However, USA firms do appear to be more cooperative during the implementation period, which, as we argue, is due to the influence of the NAO during the early crisis period between 1987-1990. The norm also appears to reverse during the immediate period following DOD exit, as USA firms are forming alliances at a more rapid rate with firms of other national headquarters, i.e. higher tie formation simultaneous with negative homophily. Conversely, JPN firms appeared to be more cooperative in general and more cooperative with each other (homophily), which speaks to prior research on the cooperative Japanese organizational form, the *Keiretsu* (Lincoln, Gerlach, and Takahashi, 1992). These findings raise interesting questions regarding the contemporary state of the high-technology landscape. For example, how might the rapid rise of non-democratic China after the turn of the 21st century effect international collaboration behavior on the global scene? Given that China has made heavy investments to stimulate technological innovation, including the creation of powerful state-owned enterprises, what are the implications for organic processes of self-organization? How might domestic network governance strategies directed at competition with an allied democracy apply in this new era?

*Limitations*: The first limitation deals with the issue of only having data on tie formation from 1987-89 (and not ongoing alliances or tie dissolution), limiting the analysis of that period to a pooled ERGM rather than the SIENA model. This pooling approach to the ERGM in the early years may discount information on temporal dynamics and may also understate the extent of alliance formation during this period. Second, there is no standard method for directly assessing whether preferential attachment is stimulated by select nodes. Rather, preferential attachment generally refers to the distribution of ties in a network. As such, we rely on proxy variables in the models. Future research could explore other methods for interacting public policy variables with structural network properties. Third, the case of Sematech and governance theory that applies to the technology sector may not generalize well to other sectors where cooperation is not necessary for innovation, or where cooperation can lead to conflicts of interest, corruption, regulatory capture, moral hazard, collusion, or anti-trust violations. Similarly, the Sematech approach may not generalize across time for different eras given varying global economic and political conditions. Such an approach may even be counterproductive where a more adversarial principal-agent relationship between government and the private sector is necessary to regulate high risk activities, such as criminal justice or warfighting. In this sense, without appropriate correcting mechanisms in place, preferential attachment may be limited as a tool for achieving public ends. Without the broader constraint of public policy to prevent corruption, preferential attachment may produce negative public outcomes, such as an unequal distribution of resources controlled by only a small group of firms. Finally, as with all non-experimental designs that lack random assignment to treatment and control groups, we cannot be certain that Sematech alone is causing increases in the probability of tie formation, or whether these increases would have been observed in the absence of Sematech. As such, we limit our conclusions to associational claims rather than using causal language. Despite these limitations we remain confident in the findings which appear to be compatible with the extant literature on the nature and effects of Sematech (e.g. Irwin and Klenow, 1996; Browning and Shetler 2000).

## 7. Conclusion

The aim of this study is to develop theory regarding how government intervention might catalyze self-organization and network evolution in strategic industries. This study shows how government support for a non-profit consortium helped to prevent market failure in a domestic industry, recover competitiveness in a global industry, and achieve increasing returns to federal investment. The integration of network governance and complexity theory suggests that network administrative organizations may accelerate self-organization in the cooperative networks of policy relevant sectors. Specifically, this study focuses on the network governance effects of a network administrative organization on the preferential attachment mechanism of the research and development network in the high-technology sector. Due to the importance of technology to the national interest, the quasi-public nature of technology, and the cooperative character of the technology sector, network-based governance interventions may provide a balanced approach for resolving complex public problems. The results of the study provide support for the theory and suggest further research is necessary to explore the dynamics of network intervention in strategic industries. Future research should further develop theory regarding the interactions between public policy interventions and mechanisms of self-organization on complex landscapes.

**Online Appendices**

Appendix A – List of Sematech Members

1987 - 1990 - AMD, AT&T, Digital, DOD/NSA, Harris, HP, IBM, Intel, LSI Logic, Micron, Motorola, NCR Corp, National, Rockwell, TI
1991 - AMD, AT&T, Digital, DOD/NSA, Harris, HP, IBM, Intel, LSI Logic, Micron, Motorola, NCR Corp, National, Rockwell, TI
1992 - AMD, AT&T, Digital, DOD/NSA, Harris, HP, IBM, Intel, Motorola, NCR Corp, National, Rockwell, TI
1993 - AMD, AT&T, Digital, DOD/NSA, HP, IBM, Intel, Motorola, NCR, National, Rockwell, TI
1994 - AMD, AT&T, Digital, DOD/NSA, HP, IBM, Intel, Motorola, National, Rockwell, TI
1995 - AMD, AT&T, Digital, DOD/NSA, HP, IBM, Intel, Motorola, National, Rockwell, TI
1996 - AMD, Digital, HP, IBM, Intel, Lucent, Motorola, National, Rockwell, TI, Hyundai, LG Semicon, Phillips, Samsung, SGS Thomson, Siemens, TSMC
1997 - AMD, Digital, HP, IBM, Intel, Lucent, Motorola, National, Rockwell, TI, Hyundai, LG Semicon, Phillips, Samsung, SGS Thomson, Siemens, TSMC
1998 - AMD, Conexant, HP, Hyundai, IBM, Intel, Lucent, Motorola, National, Philips, Siemens, STMicroelectronics, TI, TSMC
1999 - AMD, Compaq, Conexant, HP, Hyundai, IBM, Infineon, Intel, Lucent, Motorola, Philips, STMicroelectronics, TI, TSMC

Appendix B – Robustness Check SIENA Model

Table B.1 – SIENA model including 1987-93

| | 1987/1990 - 1993 |
|---|---|
| rate period 1 | 3.84*** |
| | (0.76) |
| rate period 2 | 0.81*** |
| | (0.10) |
| rate period 3 | 0.92*** |
| | (0.11) |
| Density | -2.95*** |
| | (0.27) |
| Transitivity | 0.32*** |
| | (0.08) |
| Pref. Attachment | 0.11*** |
| | (0.02) |
| USA Effect | 0.14 |
| | (0.15) |
| USA Homophily | -0.36* |
| | (0.15) |
| Japan Effect | 0.49* |
| | (0.22) |
| Japan Homophily | 0.04 |
| | (0.20) |
| Sematech Effect | 0.58** |
| | (0.21) |
| Sematech Homophily | 0.30 |
| | (0.20) |
| Firm Size Effect | 0.18* |
| | (0.09) |
| Iterations | 3191 |

***p < 0.001, **p < 0.01, *p < 0.05

Table Notes: The model includes the data from 1987-1990, which are used in the ERGM in the article. These data are pooled as the first year in the model, then the next years are 1991, 1992, 1993.

## Appendix C - ERGM & SIENA Goodness of Fit Diagnostics

The statnet program in R was used to produce the goodness-of-fit plots for the ERGM model (Figure 1) and RSiena was used of for the SIENA models (Figure 2). In Figure 1 each column corresponds to models 1 through 3 in the ERGM table. The goodness-of-fit plots demonstrate how well networks simulated from the specified model capture "out-of-model" statistics (e.g., shared partners, degree distributions, and the triad census). These out-of-model statistics are global properties of the networks that were not directly specified in the local configurations used in the ERGM and SIENA models. In the plots below, the thick black line represents a given statistics observed value. The boxplots show the simulated networks distribution based on the model parameters. The plots show that the out-of-model statistics are well captured the ERGM. For the SIENA models, the Triad Census is well fit, however, while we capture the shape of the degree distribution, we underpredict the number of actual isolates causing slight overprediction for other levels of the degree distribution.

Appendix C, **Figure C.1 – ERGM goodness-of-fit diagnostics**

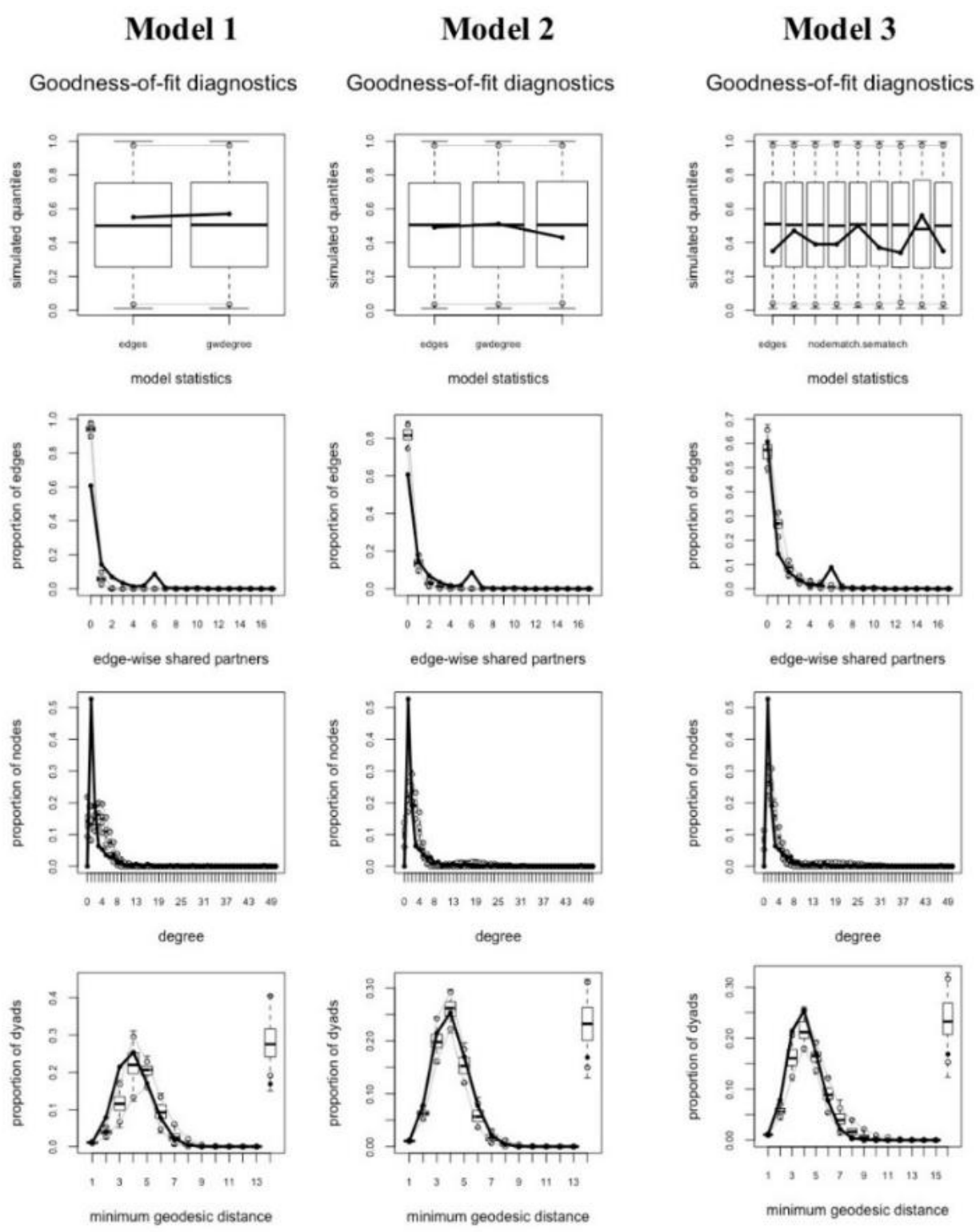

Note: The goodness-of-fit diagnostic plots depict the observed network (sold black line) in relation to the simulated network (thin gray lines) with 95% confidence intervals.

Appendix C, **Figure C.2 – SIENA goodness-of-fit diagnostics**

1991:1993

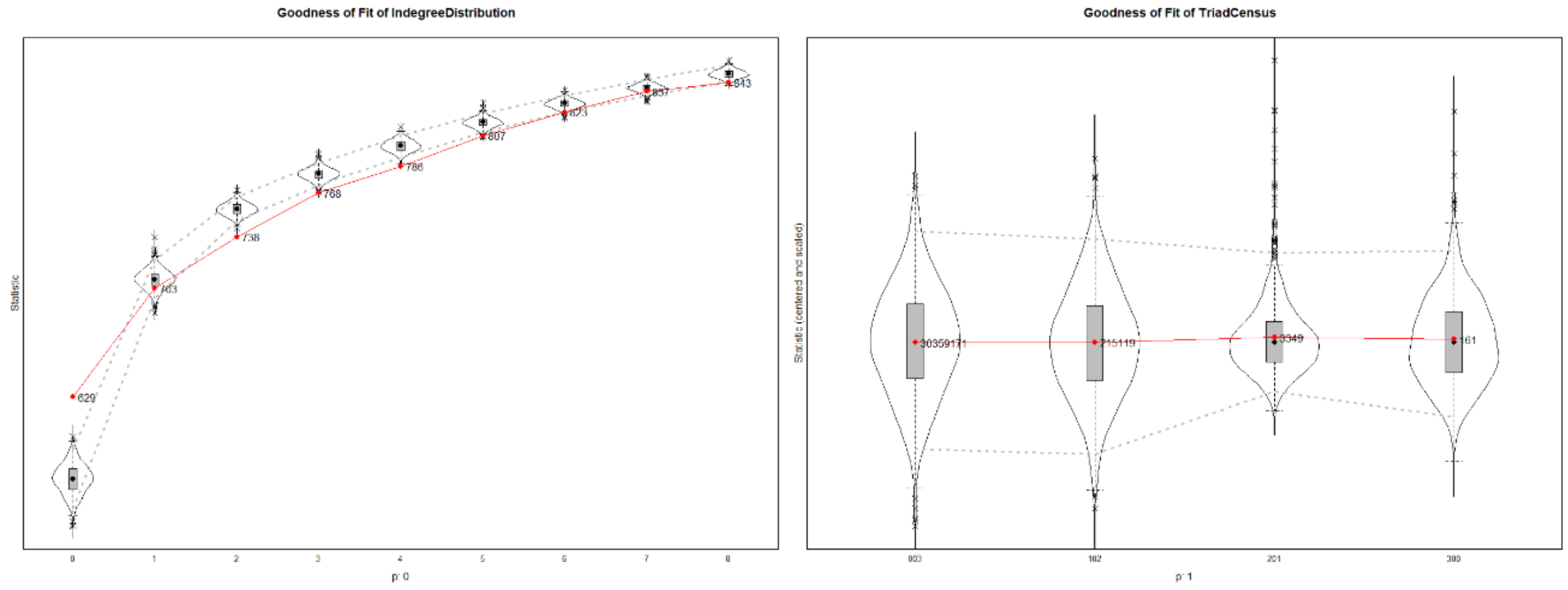

1993:1995

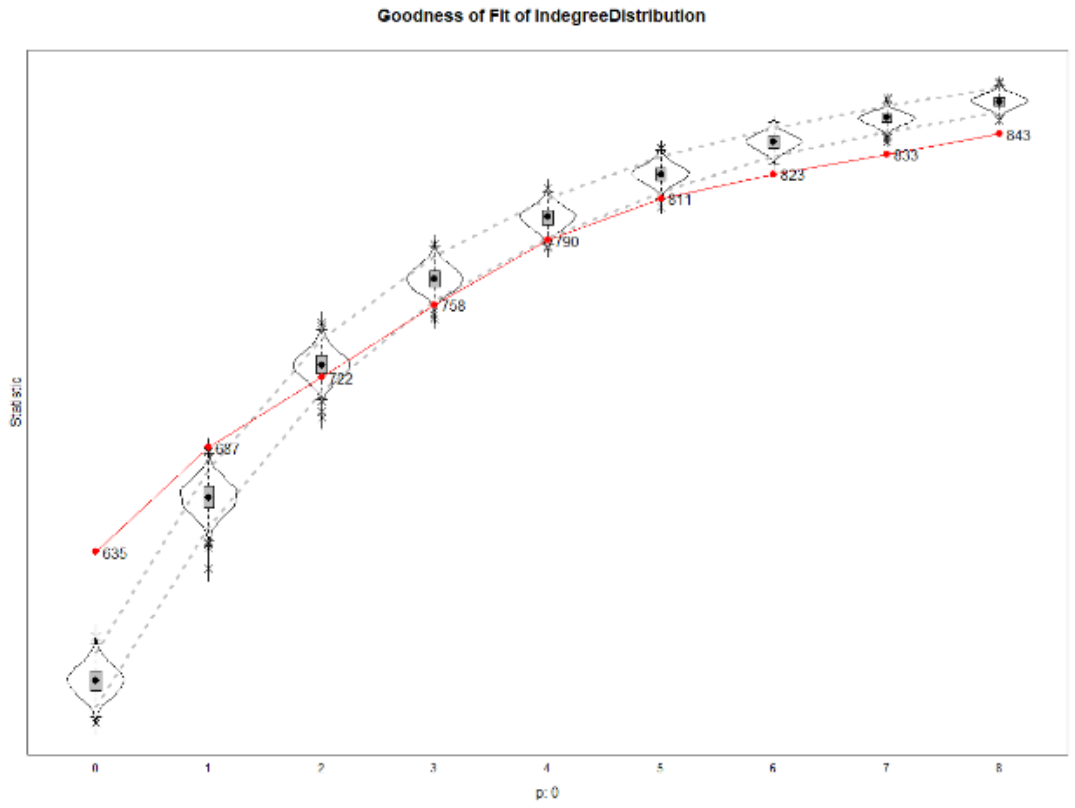

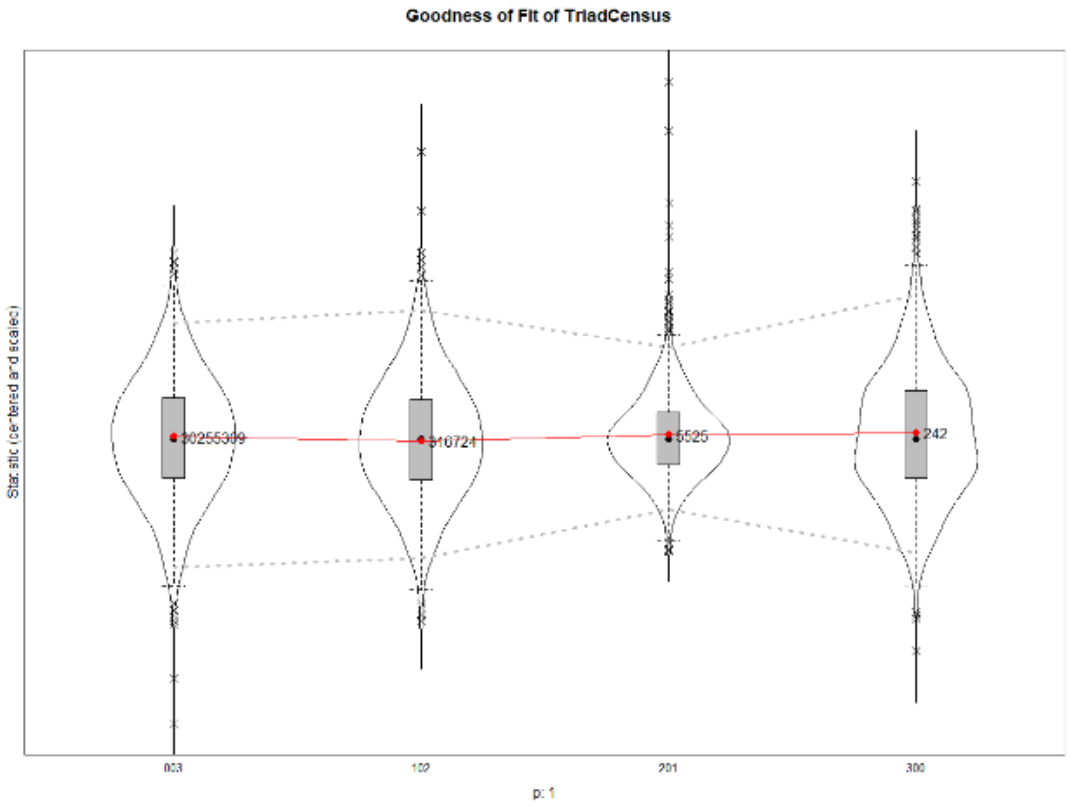

1995:1997

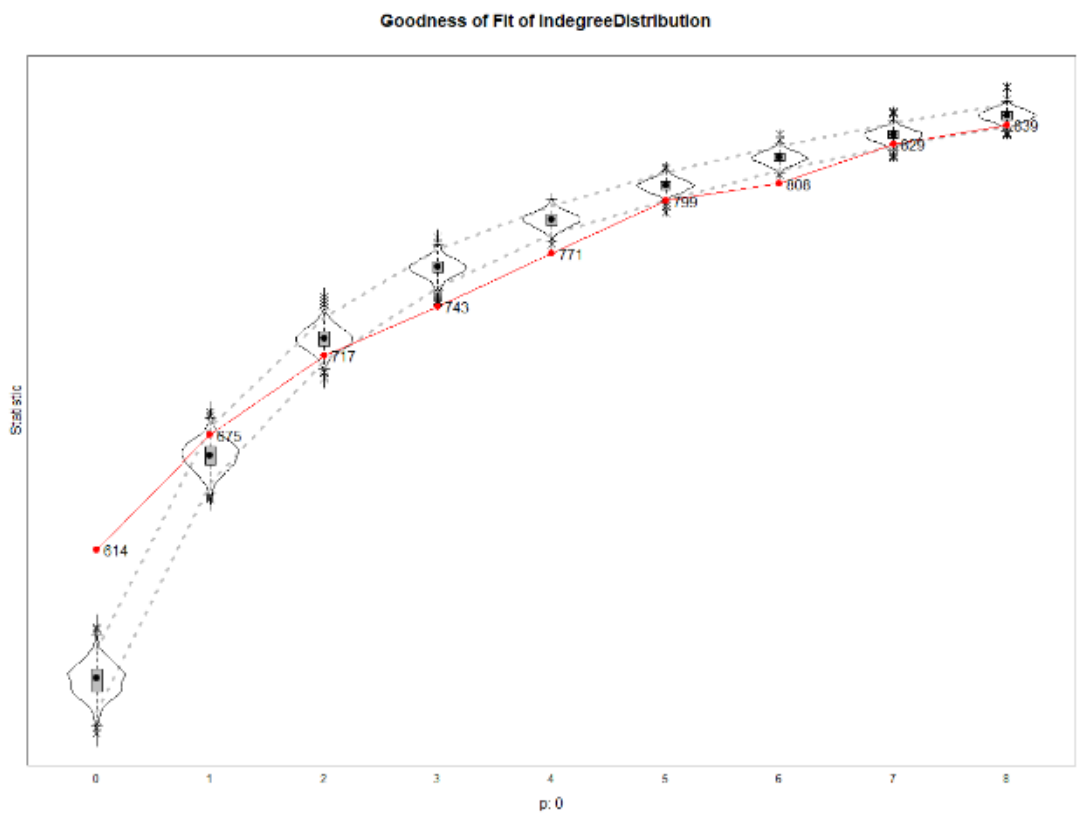

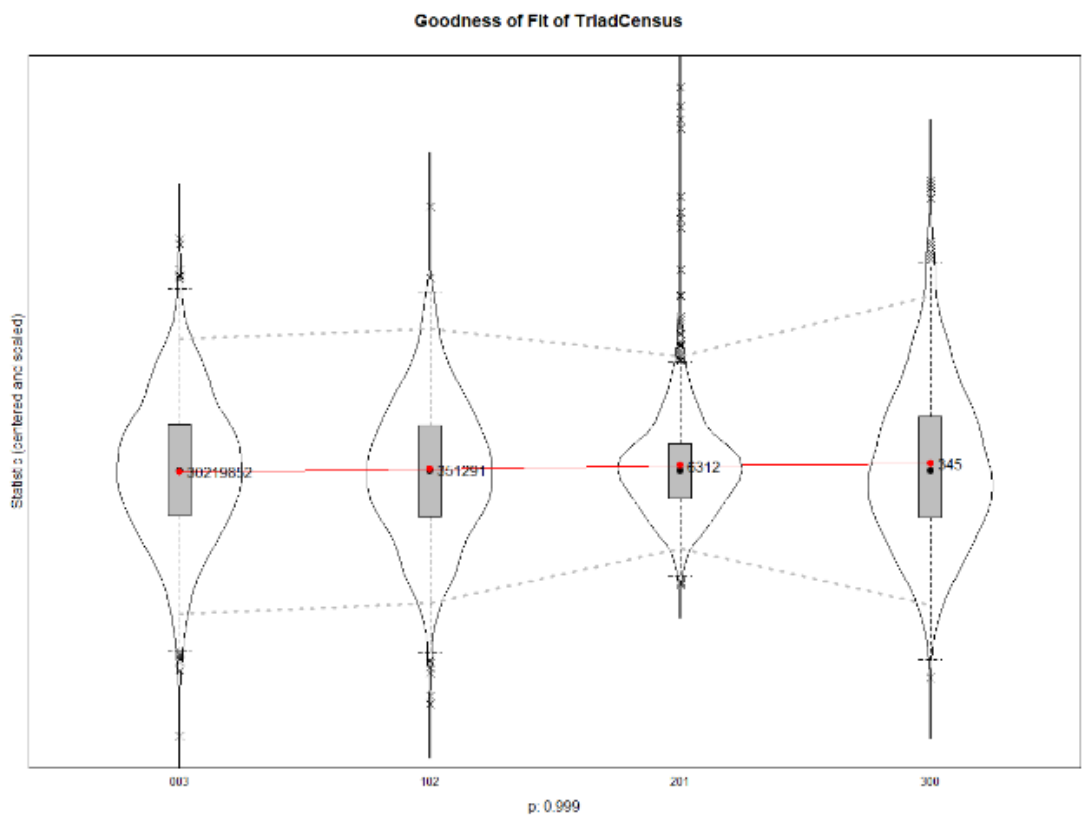

1997:1999

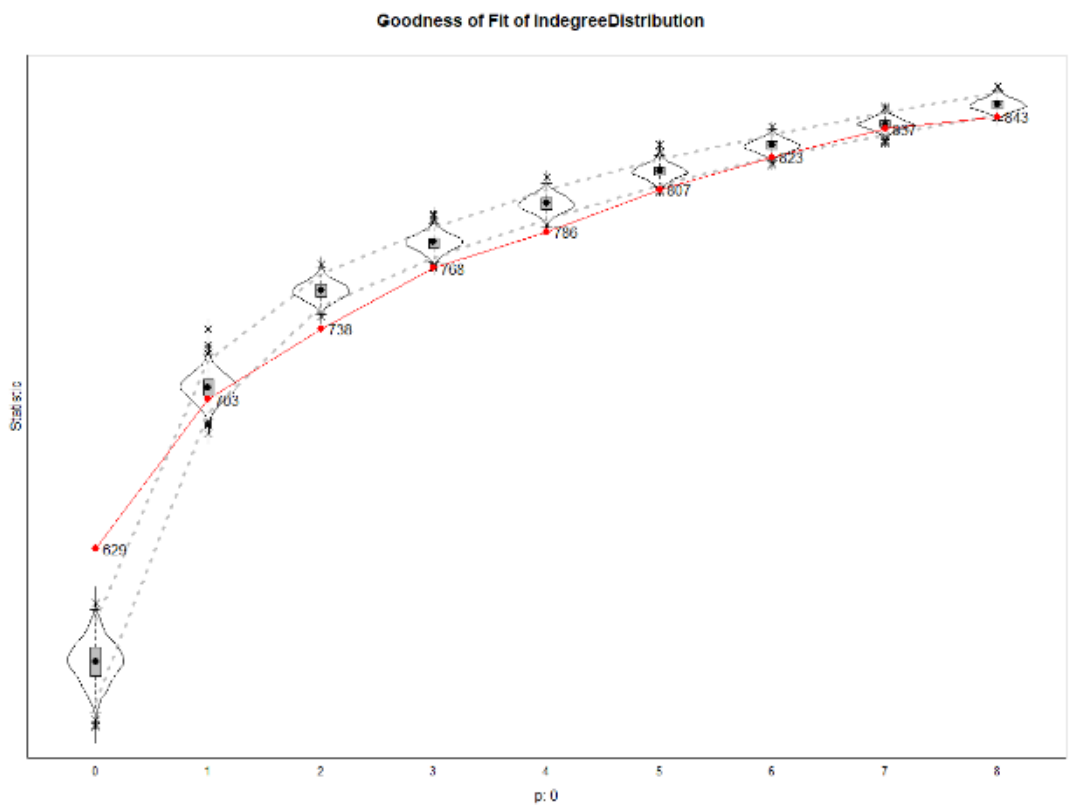

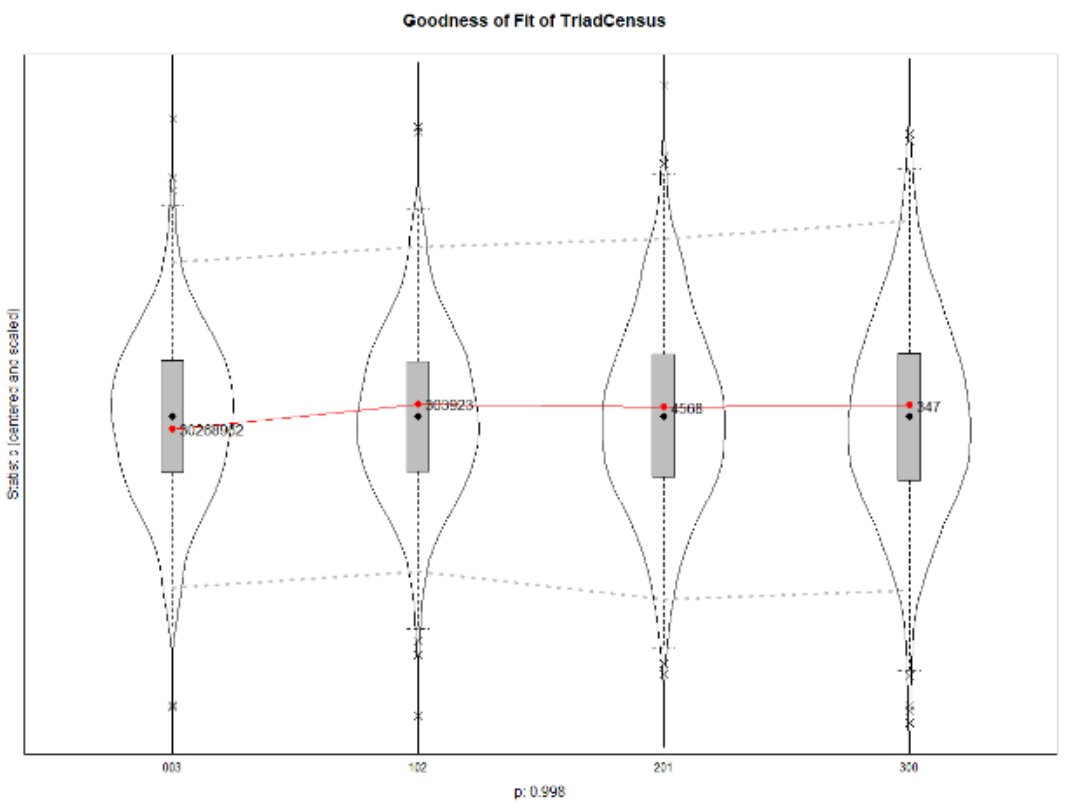